\documentclass{llncs}
\usepackage{graphicx}
\usepackage{multirow}
\usepackage[caption=false]{subfig}
\usepackage{nicefrac}
\usepackage[binary-units=true]{siunitx}
\DeclareSIUnit \dBm {dBm}
\DeclareSIUnit \dB {dB} 
\DeclareSIUnit \dBi {dBi} 
\DeclareSIUnit \Kbps {Kbps}
\DeclareSIUnit \Mbps {Mbps}
\DeclareSIUnit \Gbps {Gbps}
\DeclareSIUnit \kBps {kBps}
\DeclareSIUnit \MBps {MBps}
\DeclareSIUnit \GBps {GBps}
\usepackage{hyperref}
\DeclareSIUnit{\million}{\text{million}}

\newcolumntype{C}{ >{\centering\arraybackslash} m{2.3cm} }
\begin{document}
%
\title{A City-Scale ITS-G5 Network for Next-Generation Intelligent Transportation Systems: Design Insights and Challenges}
%
%

\author{Ioannis Mavromatis \and Andrea Tassi \and Robert J. Piechocki \and Andrew Nix}

%
\institute{Department of Electrical and Electronic Engineering, University of Bristol, UK\\
Emails: \email{\{ioan.mavromatis, a.tassi, r.j.piechocki, andy.nix\}@bristol.ac.uk}}
\maketitle              
\begin{abstract}
As we move towards autonomous vehicles, a reliable Vehicle-to-Everything (V2X) communication framework becomes of paramount importance. In this paper we present the development and the performance evaluation of a real-world vehicular networking testbed. Our testbed, deployed in the heart of the City of Bristol, UK, is able to exchange sensor data in a V2X manner. We will describe the testbed architecture and its operational modes. Then, we will provide some insight pertaining the firmware operating on the network devices. The system performance has been evaluated under a series of large-scale field trials, which have proven how our solution represents a low-cost high-quality framework for V2X communications. Our system managed to achieve high packet delivery ratios under different scenarios (urban, rural, highway) and for different locations around the city. We have also identified the instability of the packet transmission rate while using single-core devices, and we present some future directions that will address that.
\end{abstract}

\section{Introduction}
The Automotive Industry is progressively commercialising several advanced features such as lane-keeping assistance, forward collision braking, etc. Even the most pessimistic market analysis envisage that fully autonomous vehicles will flood the global market by 2025~\cite{projection}. Autonomous vehicles are expected to be equipped with several sensors that will assist their autonomous functionalities~\cite{fully_autonomy}. However, the most critical enabler of the full autonomy will be the communication framework~\cite{connectivity} among the vehicles, i.e. Vehicle-to-Vehicle (V2V), and between the vehicles and the infrastructure network, i.e. Vehicle-to-Infrastructure (V2I).

The communication framework is essential as the information exchanged can increase the vehicle safety, provide new services, reduce traffic jams improving the fleet routing, etc. For these reasons, \SI{75}{\mega\hertz} of the spectrum have been allocated in the \SI{5.9}{\giga\hertz} band for Dedicated Short Range Communication (DSRC) to be used for the Cooperative Intelligent Transportation Systems (C-ITSs). The DSRC radio technology was standardized by IEEE in the 802.11p standard~\cite{ieee80211pStandard}, describing the PHY and the MAC layer of the framework, as well as in~\cite{std16093} and~\cite{std16094} describing the networking services and the multi-channel operation respectively. 

Performance enhancement of DSRC communications is a hot research topic. The development of a robust Vehicle-to-Everything (V2X) communication framework, able to guarantee the exchange of information between the Connected and Autonomous Vehicles (CAVs), remains a challenge. Most of the research activities on Vehicular Ad-Hoc Networks (VANETs) and the DSRC focus on computer simulations and theoretical models (e.g.~\cite{example1,example2}). The importance of simulations, as well as their limitations, were discussed in~\cite{agileCalibration} showing that, in larger scale scenarios, the existing simulation frameworks lack in accuracy and realism. To that extent, we will focus our research on building and deploying a real-world large-scale testbed for V2I and V2V communications that could offer:
\begin{itemize} 
	\item Continuous availability for delay-critical applications.
    \item Full-stack system implementation to support various vehicular applications.
    \item Centralised coordination via a Software-Defined Networking (SDN)-like framework.
    \item Open-source operating system for customisability and compatibility with Fog and Cloud Computing architectures~\cite{5683765}.
    \item Reduced cost for large-scale deployments.
\end{itemize}
Our experimental testbed is currently deployed in the City of Bristol, UK. Similar activities can be found in the literature (e.g.~\cite{CarTel,Cabernet,harbornet}), however, only~\cite{harbornet} considered a V2V communication framework. The systems mentioned above mainly rely on Commercial Off-the-Self (COTS) implementations and license-based products. Our prototyped system relies on open-source firmware and low-cost hardware components. Besides, to the best of our knowledge, none of the existing works considered integration with the Fog Computing paradigm. Finally, throughout our three days of field trials, we logged the messages that we generated and exchanged in a V2I and V2I fashion. Our dataset is freely available and can be downloaded from~\cite{dataset}. Later in this work, we will further explain our experimental setup and the messages exchanged.

This paper is organized as follows. In Sec.~\ref{sec:architecture} we present our testbed architecture and describe our prototyped setup in terms of the hardware and the software used. Sec.~\ref{sec:trials} describes the testbed deployed around the City of Bristol, UK as well as the field trials that we conducted using this experimental setup. This section also introduces the initial performance investigation of our system. Based on the knowledge acquired from the aforementioned field trials, we later identify the drawbacks that should be addressed in the future. Finally, in Sec.~\ref{sec:conclusions} we summarise our key findings, we comment on the knowledge acquired from this real-world experimentation, and we introduce some ideas for future research.

\section{Experimental Testbed Architecture}\label{sec:architecture}

\subsection{Description of the System Architecture}
\begin{figure}[t]     
\centering
\includegraphics[width=\textwidth]{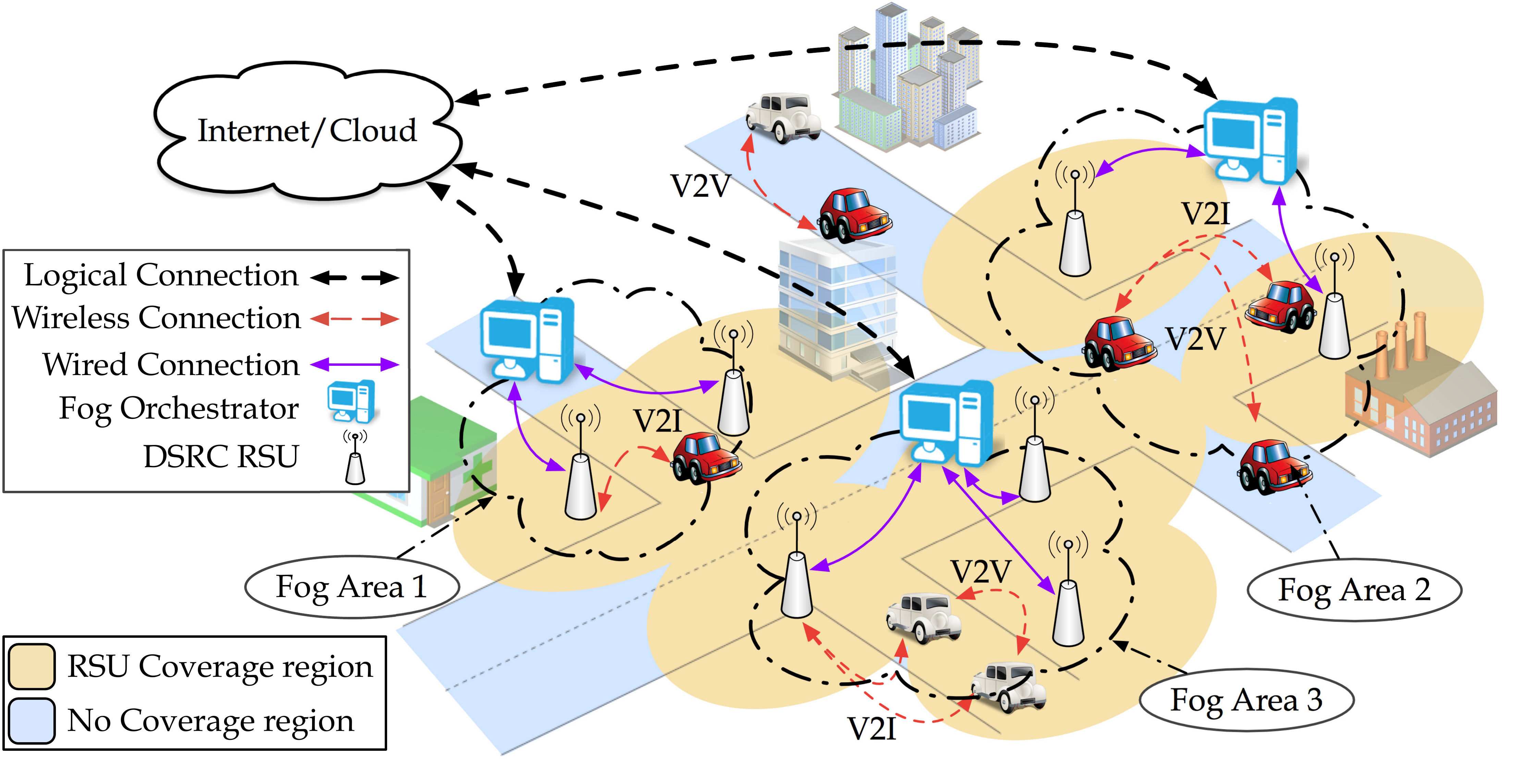}
    \caption{A general overview of the considered system model. The C-ITS design framework ensures V2X connectivity and a NFV architecture in the infrastructure domain.}
    \label{fig:network}
\end{figure}

Our developed experimental VANET testbed consists of different devices and entities. Each one will play a significant role in the operation of our system. An ideal design paradigm can be found in Fig.~\ref{fig:network}. The different devices that will form our system paradigm are the following:
\begin{itemize}
	\item \emph{Road Side Units (RSUs):} Network infrastructure devices, mounted on several building, and connected to a centralised control plane to provide V2I connectivity.
    \item \emph{On-Board Units (OBUs):} Devices installed in the vehicles, able to exchange safety critical messages with the RSUs and other vehicles.
    \item \emph{Fog Orchestrators (FOs):} Devices that centrally manage the different clustered management areas, called \emph{Fog Areas}, ideally with one-hop distance from the RSUs to reduce the end-to-end delay.
\end{itemize}

As shown in Fig.~\ref{fig:network} the RSUs and OBUs will be connected to each other using IEEE 802.11p/DSRC links. In our system, vehicles can connect to another vehicle (when driving at no coverage regions) or to RSUs (when within the RSU coverage range). In our system paradigm, we assume that our infrastructure network is clustered in different management areas called \emph{Fog Areas}. FOs manage each Fog Area and share a wired connection with the different RSUs. Being one-hop away from the RSUs, they can be used to process all the time-critical information received or generated at the infrastructure side with reduced end-to-end delay. Finally, our system will interact with a cloud-based city-wide connection, interfacing with the different FOs. The cloud-based service will be responsible for recording city-scale data, interconnecting the different FOs and Fog Areas and pushing city-scale policies in the entire network. Some more details about this system architecture can be found in~\cite{architecture,tvtAndrea,vtcIoannis,iccIoannis}. In the next section, we will describe in greater detail the testbed components that we have already designed and deployed around the City of Bristol. Compared to our work in~\cite{agileCalibration}, in this work, we will focus more on the exchange of Cooperative Awareness Messages (CAMs) on V2I and V2V links as well as the challenges that we faced concerning the large-scale deployment and our solutions for them. Our discussion on the current large-scale deployment will be followed by some preliminary results from our experimental study. Finally, for our current work, the idea of Fog Areas and the deployment of FOs was not considered. This will be a task for our future research activities.

\subsection{Description of the Experimental Setup}\label{sub:devices}

\begin{figure}[t]
\begin{minipage}{.28\linewidth}
\centering
\subfloat[IEEE 802.11p / DSRC RSU units.]{\label{fig:units}\includegraphics[width=0.98\linewidth]{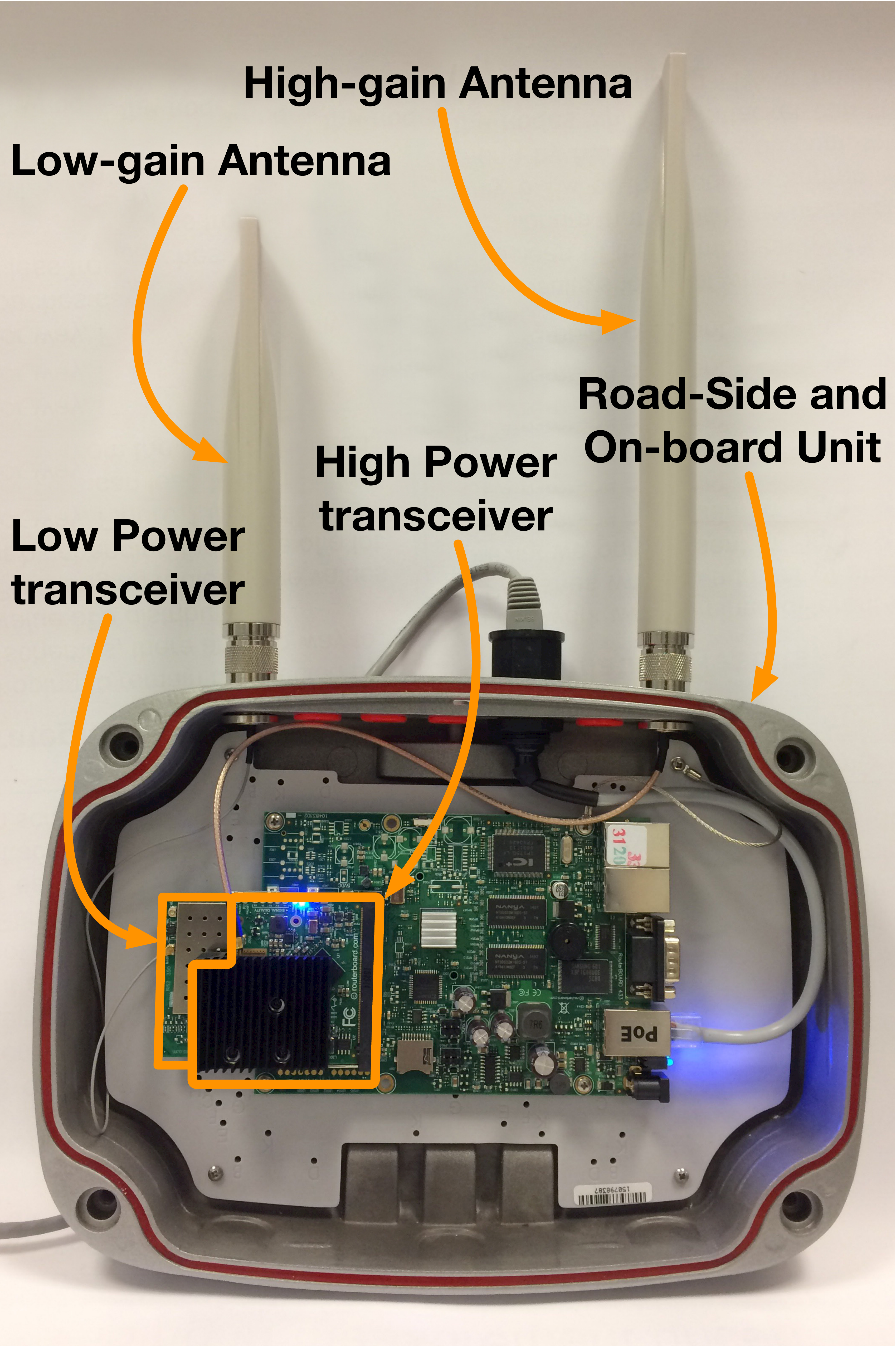}}
\end{minipage}%
\begin{minipage}{.72\linewidth}
\centering
\subfloat[IEEE 802.11p / DSRC OBU units.]{\label{fig:units2}\includegraphics[width=0.98\linewidth]{{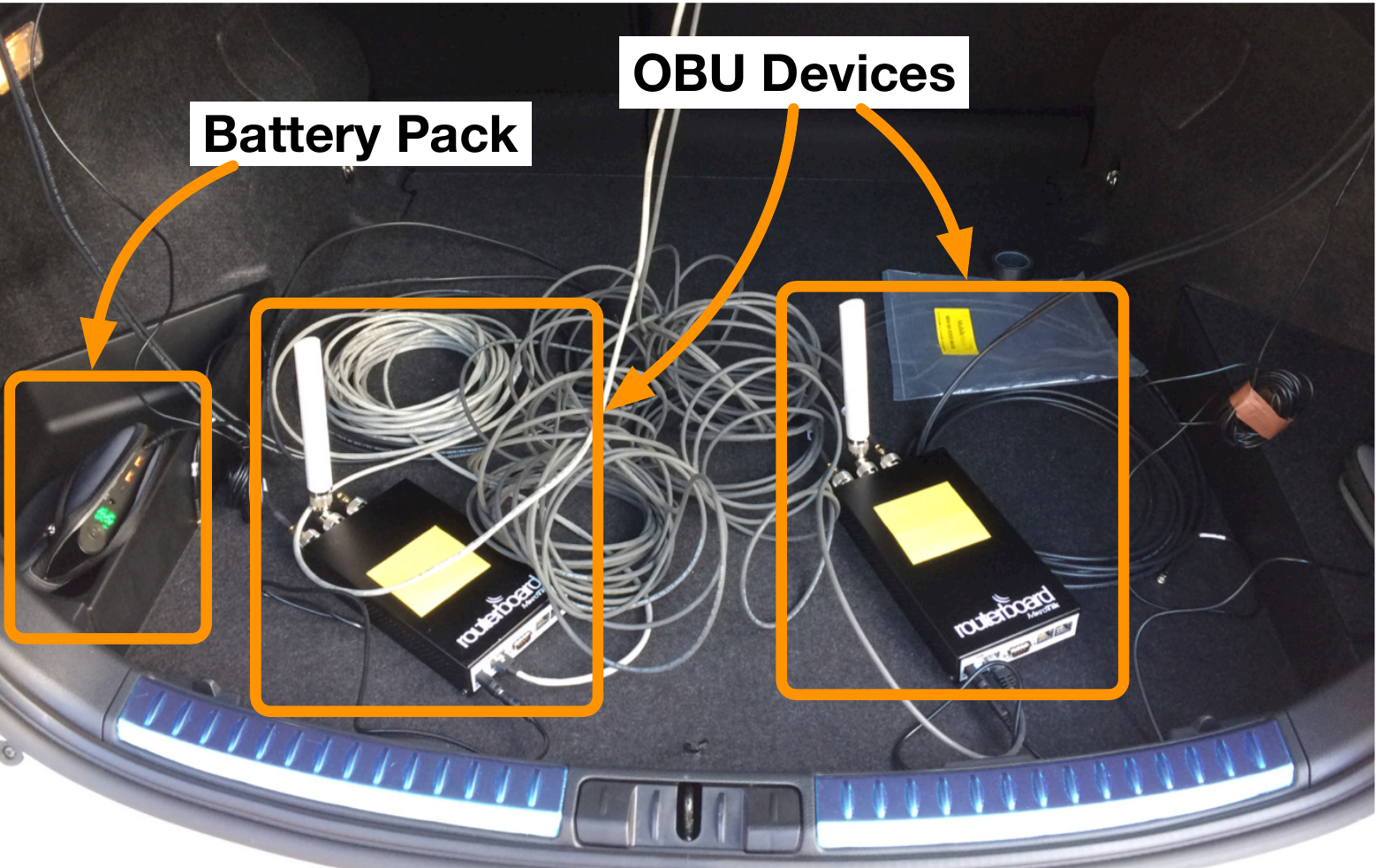}}}
\end{minipage}\par\medskip
\centering
\subfloat[OBU antenna mounted on the roof of the car.]{\label{fig:car}\includegraphics[width=0.7\linewidth]{{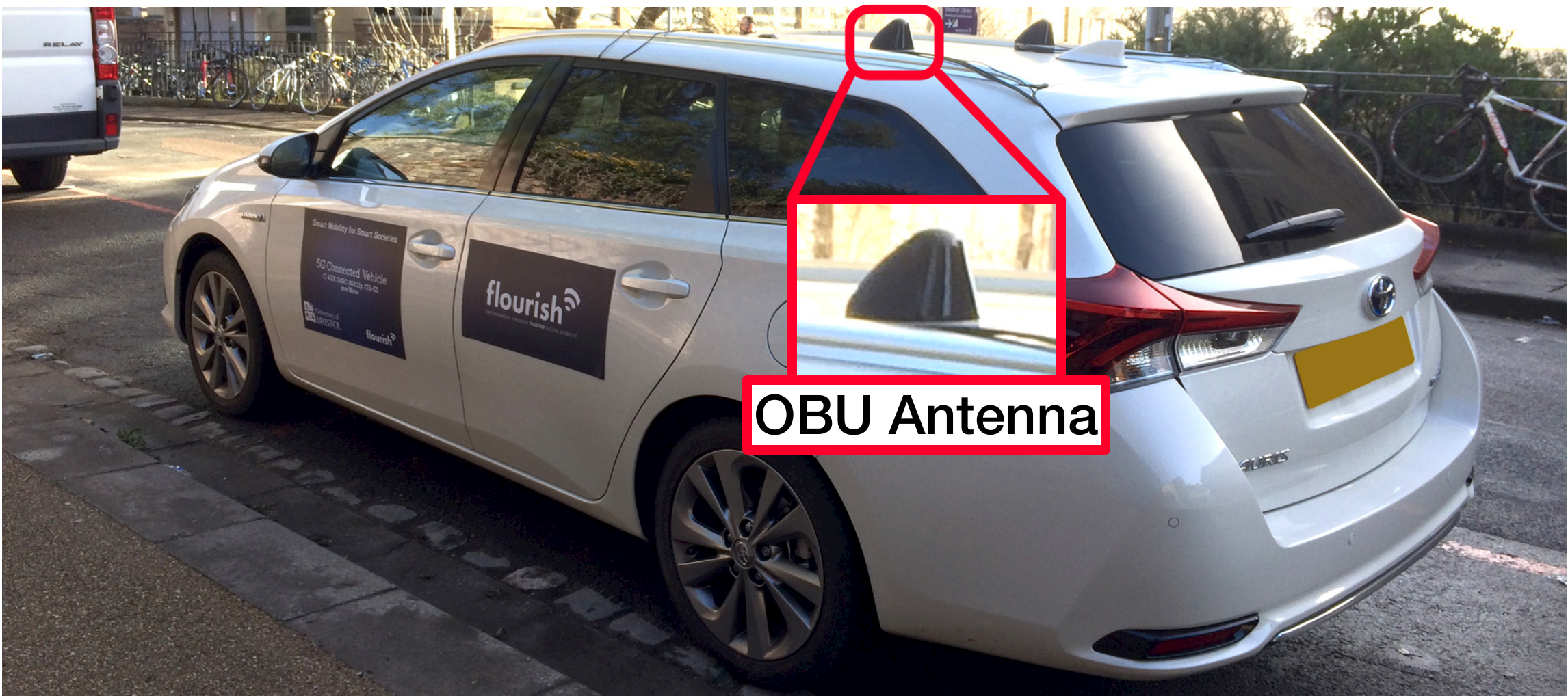}}}
\caption{Our experimental setup. We prototyped both RSUs and OBUs units, equipped them with different antennas and conducted our trials around the City of Bristol.}
\label{fig:unitsAll}
\end{figure}

For our experimental validation, we prototyped an open-source IEEE 802.11p/DSRC testbed (Fig.~\ref{fig:unitsAll}). Our devices, under ideal-like Line-of-Sight (LOS) conditions, were able to achieve good performance and high Packet Delivery Rate (PDR) for distances up to \SI{700}{\meter} (as proven in~\cite{agileCalibration}).

The devices were designed to be used as both RSUs and OBUs (Figs.~\ref{fig:units} and~\ref{fig:units2}). They were equipped with a Mikrotik RB433 single-board computer (CPU \SI{300}{\mega\hertz}, \SI{64}{\mega\byte} RAM, \SI{64}{\mega\byte} storage space, x3 Ethernets, x3 MiniPCI slots)~\cite{rb433}. Also, two wireless IEEE 802.11a NICs were used for redundancy, one regarded as High Power (HP) and the second one regarded as Low Power (LP). The wireless interfaces of the RSUs and the OBUs in our system are accompanied by different antennas as shown in Figs.~\ref{fig:units} and~\ref{fig:car}, one bolted on the RSUs and the second magnetically attached to the roof of our vehicles. Our RSU devices were powered up via Power-over-Ethernet (PoE), while a battery pack was used for the OBUs to avoid the voltage spikes experienced when using a lighter inverter within the vehicle. All the device and the key driver characteristics can be found in Table~\ref{table:characteristics}.

\begin{table}[t]
\renewcommand{\arraystretch}{1.3}
\centering
    \begin{tabular}[b]{r|C|C|C|C}
                 &  LP-RSU             & LP-OBU             &  HP-RSU & HP-OBU                             \\ \hline \hline
    Model        &  \multicolumn{2}{c|}{Mikrotik R52H~\cite{r52h}}      &  \multicolumn{2}{c}{Mikrotik R5SHPn~\cite{R5SHPn}}         \\ \hline
    TX Power     &  \multicolumn{2}{c|}{\SI{25}{\dBm}}      &  \multicolumn{2}{c}{\SI{29}{\dBm}}           \\ \hline
    Antenna Gain &  \SI{7}{\dBi}  & \SI{5}{\dBi}  &  \SI{9}{\dBi}   &     \SI{5}{\dBi} \\ \hline
    Linux Driver &  \multicolumn{2}{c|}{ath5k}              &  \multicolumn{2}{c}{ath9k}                   \\ \hline
    Bandwidth    &  \multicolumn{4}{c}{\SI{10}{\mega\hertz}}                                              \\ \hline
    Frequency   & \multicolumn{2}{c|}{\SI{5.89}{\giga\hertz}}  & \multicolumn{2}{c}{\SI{5.9}{\giga\hertz}}  \\ \hline
    $CW_{\mathrm{min}}, CW_{\mathrm{max}}$ &  \multicolumn{4}{c}{$\left[15, 1023\right]$} \\ \hline
    MCS    &  \multicolumn{4}{c}{QPSK \nicefrac{1}{2}} \\ \hline
    \end{tabular}
    \vspace{2mm}
    \caption{Wireless Network Interface Controller Characteristics}
	\label{table:characteristics}
\end{table}    

\begin{figure}[t]     
\centering
\includegraphics[width=1\textwidth]{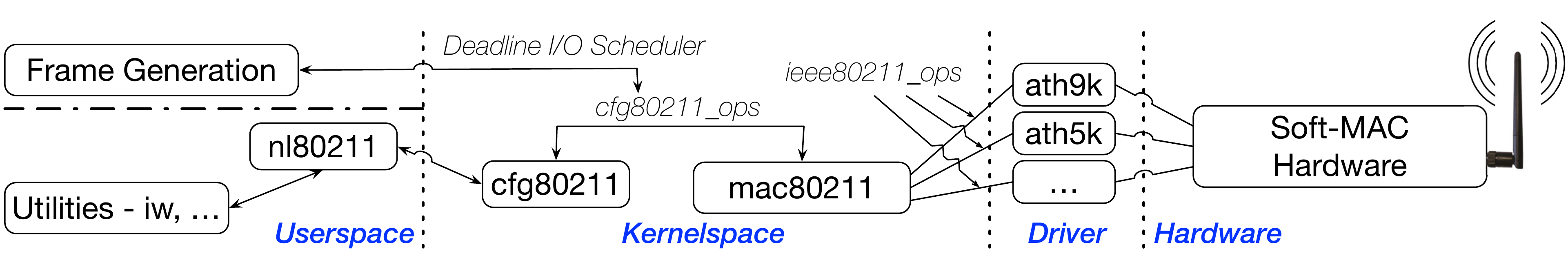}
    \caption{Linux Kernel Modules modified to enable the IEEE 802.11p/DSRC capabilities in our system.}
    \label{fig:drivers}
\end{figure}

OpenWRT\footnote{OpenWRT Barrier Breaker Release no. 14.07 - https://openwrt.org/}, a low-latency Linux distribution, was used as the operating system for both devices. Both drivers (Table~\ref{table:characteristics}) and the Linux kernel modules were modified accordingly to enable IEEE 802.11p compatibility (Fig.~\ref{fig:drivers}). The \SI{5.9}{\giga\hertz} band was added to the regulatory domain and the Outside the Context of a BSS (OCB) mode was enabled in the MAC layer, to allow NICs to operate without being associated. The values for the contention windows and the Modulation and Coding Rates (MCSs) were chosen to follow the regulation for the ITS-G5 standard specification. Integration with a GPS dongle via a USB interface was enabled. A beaconing interface was also developed that generates IEEE 802.11p DSRC CAMs and broadcasts them in the network. More details about the modifications can be found in~\cite{agileCalibration}. 

\begin{figure}[t]     
\centering
\includegraphics[width=\textwidth]{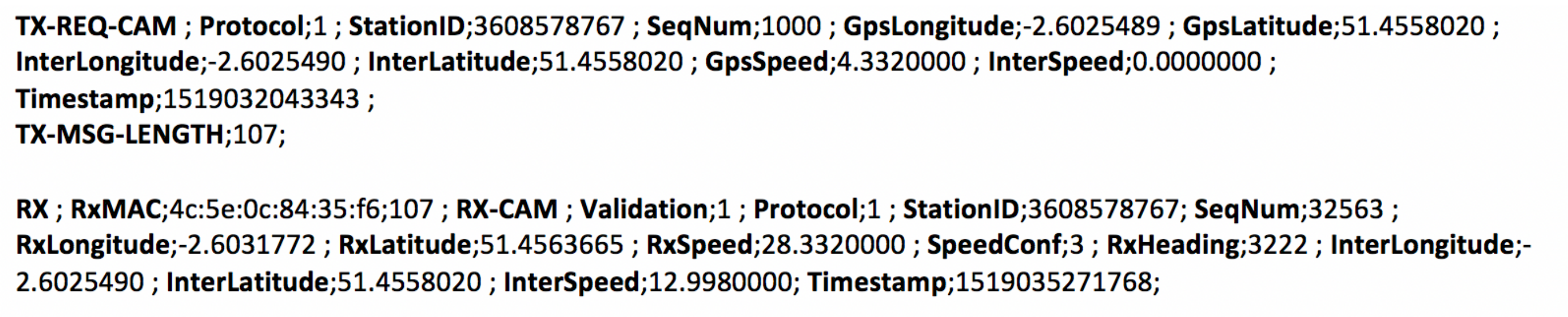}
    \caption{Example of the log file generated at the transmitter and the receiver side.}
    \label{fig:logExample}
\end{figure}

The GPS coordinates, the speed, the heading and the timestamp of the GPS are being encapsulated within the transmitted CAMs. A logging interface was designed that logs all the packets generated, transmitted and received. An example of the packets exchanged can be found in Fig.~\ref{fig:logExample}. At the TX side, the acquired GPS coordinates are represented as \emph{GpsLongitude}, \emph{GpsLatitude}, being respectively the longitude and latitude values. The \emph{InterLongitude} and \emph{InterLatitude} values are the interpolated values based on the acquired GPS coordinates. The \emph{SeqNum} is the sequence number of the packet generated (starting at zero when the device boots up). The \emph{GpsSpeed} and \emph{InterSpeed} are the acquired values from the GPS dongle and the interpolated value respectively. Finally, the \emph{Timestamp} is the time that the packet is generated, given in Unix Epoch format. The rest of the fields are used for debugging purposes only.

At the RX side, the \emph{RxMAC} is logged at first, which is the MAC address of the device transmitted the packet. \emph{RxLongitude} and \emph{RxLatitude} are the GPS coordinates encapsulated in the transmitted packet. Finally, the \emph{InterLongitude} and \emph{InterLatitude} values represent the current longitude and latitude of the receiver, acquired from the GPS dongle and interpolated later. The remaining values are similar to the transmitted packet. The above system is highly customisable, and in the future, more features extracted from different sensors can be encapsulated in the exchanged frames to introduce different vehicular applications and expand the cooperative awareness of a vehicle.

\section{Field Trials and Preliminary Results}\label{sec:trials}

\begin{figure}[t] 
	\centering
	\subfloat[Both Vehicles within RSU coverage - HP transceiver.\label{fig:heatmapResultsHP1}]
        {\includegraphics[width=0.48\textwidth]{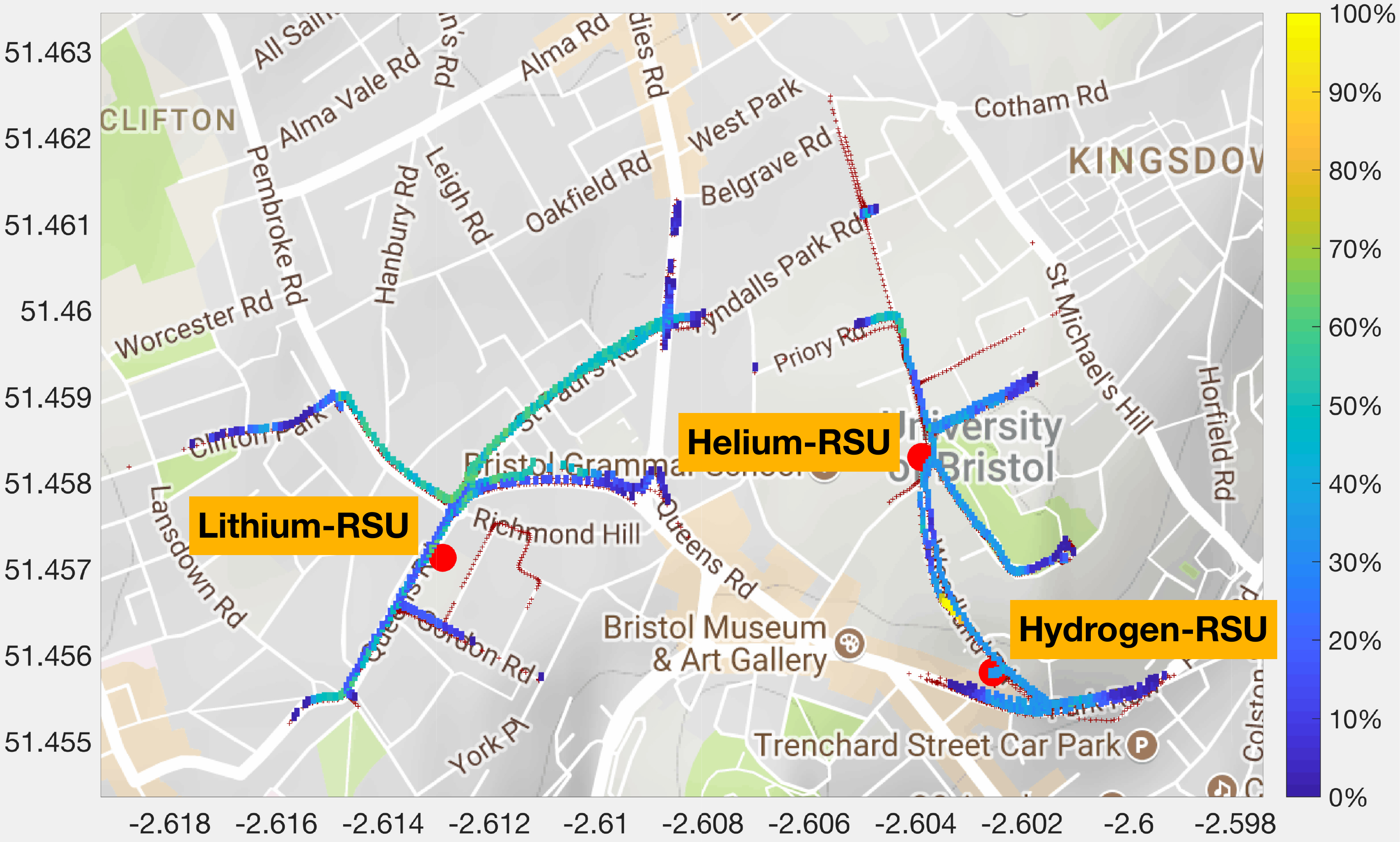}}
    \hfill
	\subfloat[Both Vehicles within RSU coverage - LP transceiver.\label{fig:heatmapResultsLP1}]
         {\includegraphics[width=0.48\textwidth]{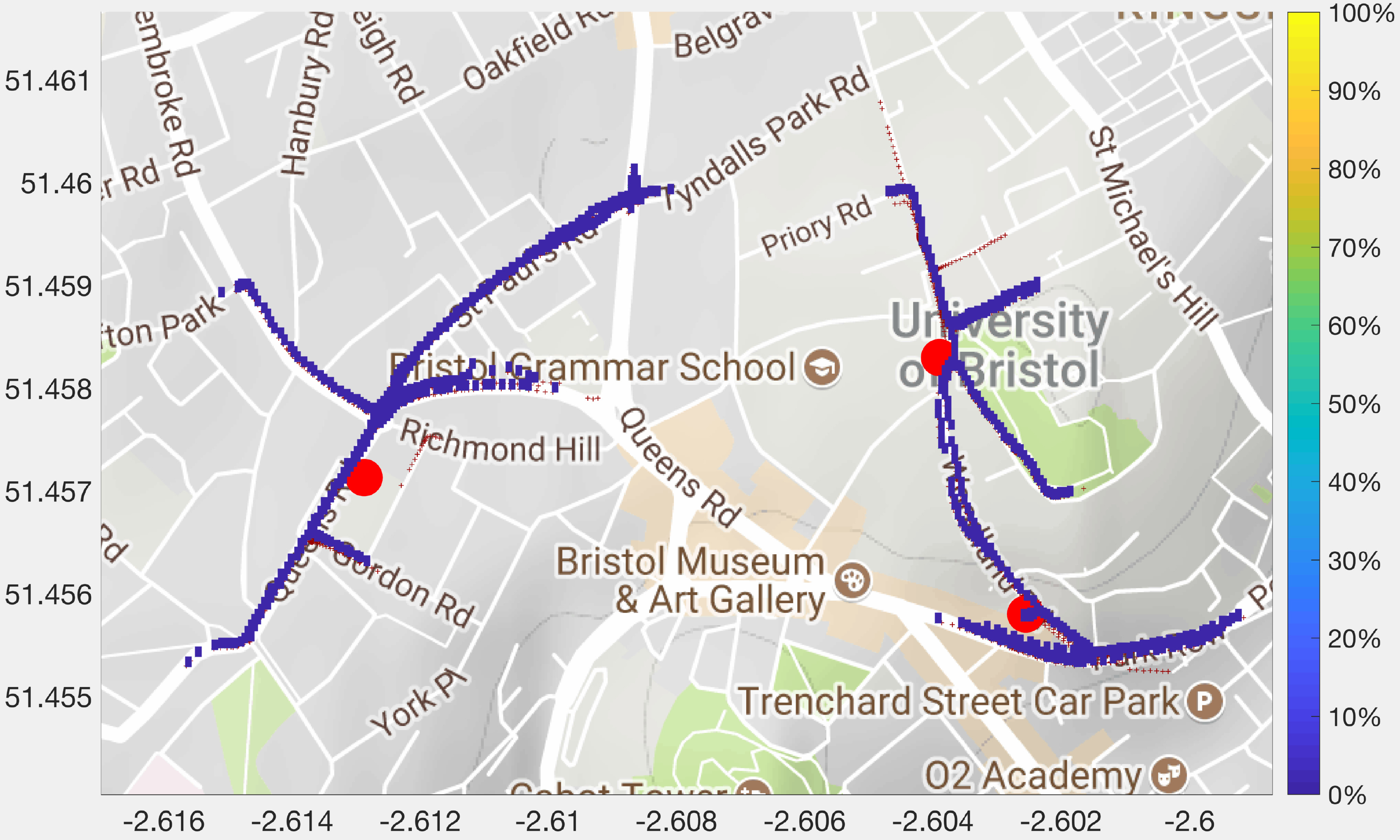}}

	\subfloat[Only Vehicle 1 within RSU coverage - HP transceiver.\label{fig:heatmapResultsHP2}]
         {\includegraphics[width=0.48\textwidth]{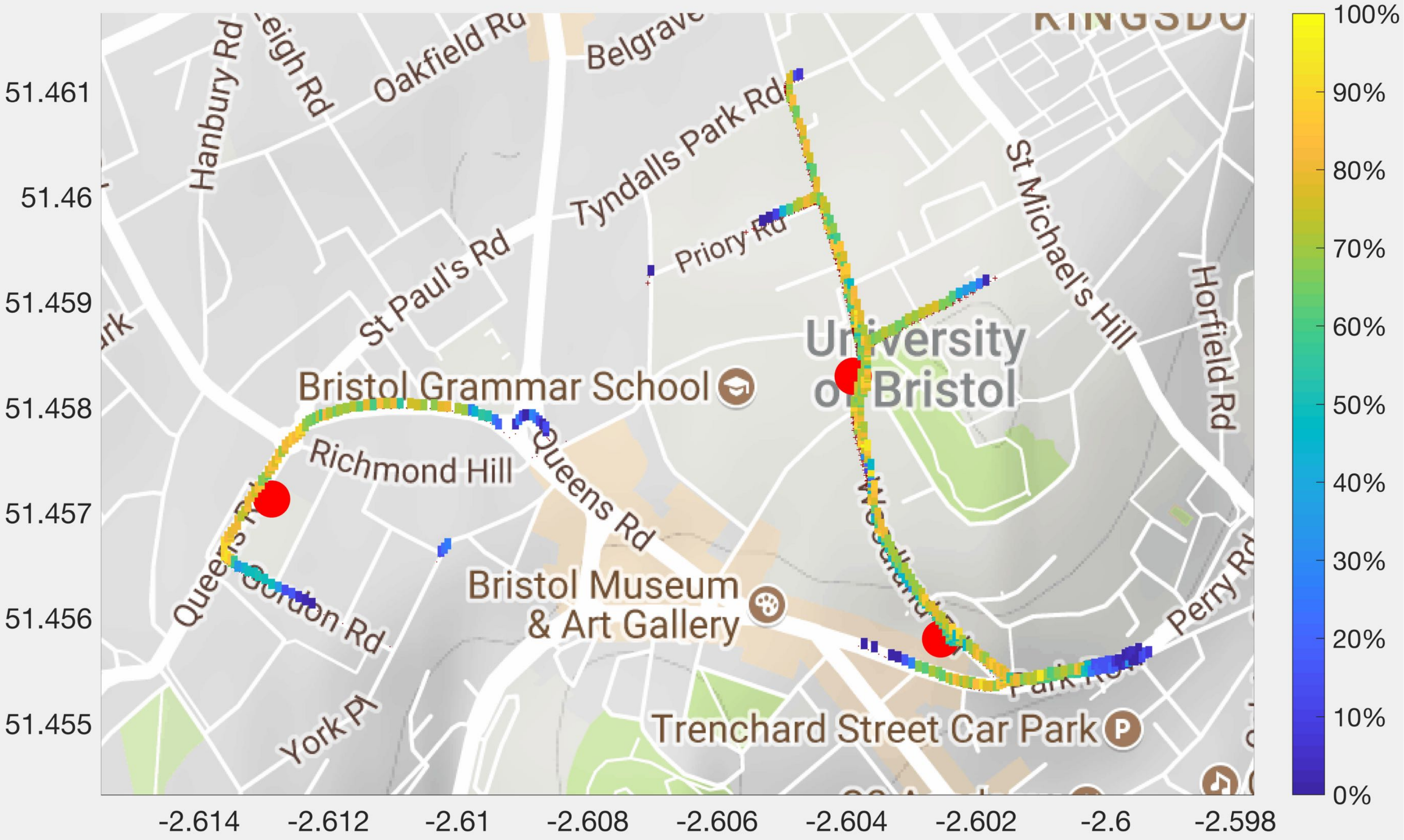}}
    \hfill
	\subfloat[Only Vehicle 1 within RSU coverage - LP transceiver.\label{fig:heatmapResultsLP2}]
        {\includegraphics[width=0.48\textwidth]{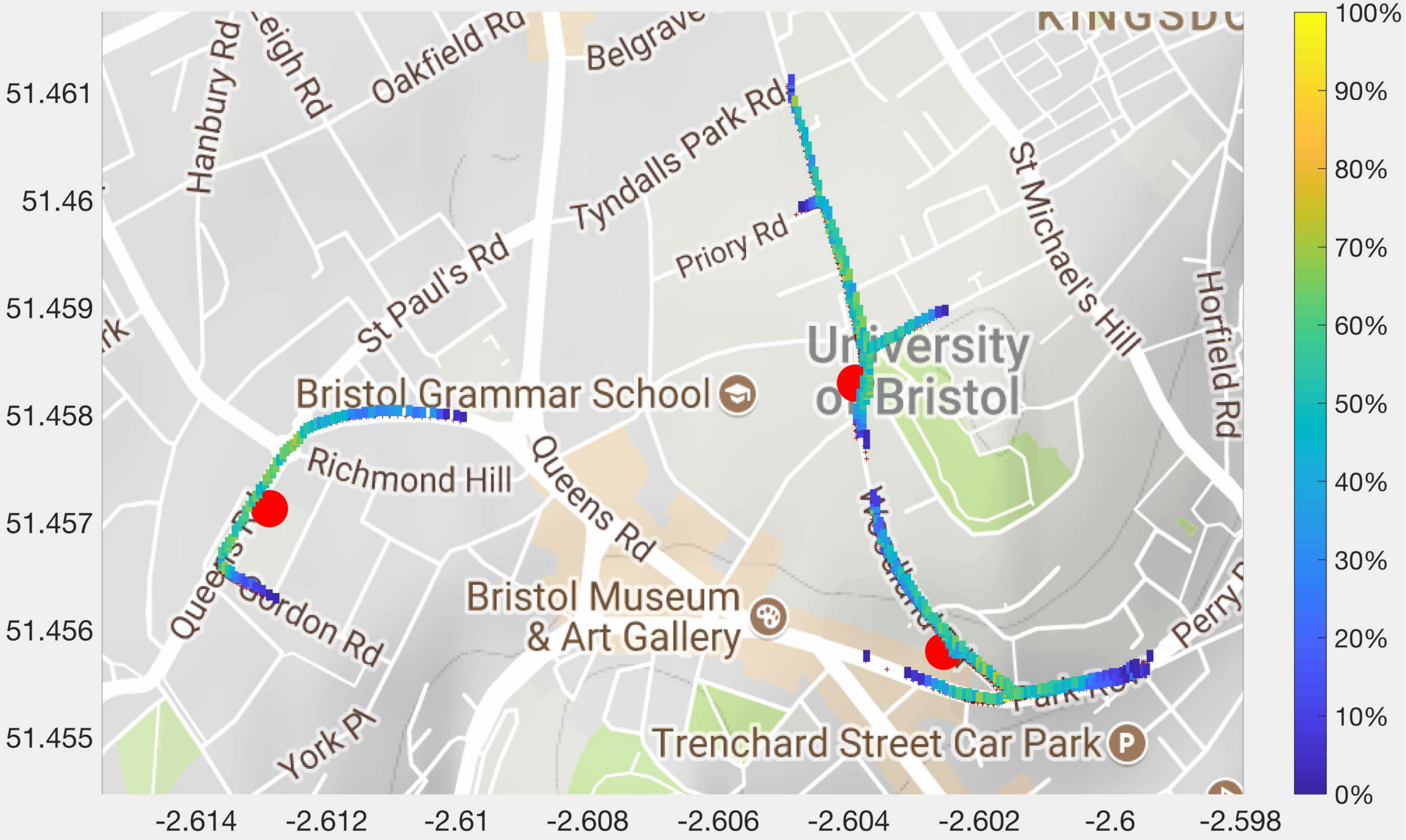}}
	\caption{Heatmap results for different V2I scenarios (HP and LP NICs).}
    \label{fig:heatmapResults}
\end{figure}

The testbed mentioned above was evaluated under a city-scale deployment during three days of field trials. Throughout the entire evaluation process, we tested various vehicular communication scenarios (both V2V and V2I) under various conditions (urban, rural, highway). The idea behind these field trials was to test the performance of our devices, identify the limitations of our system and find ways to overcome them, and finally get a more in-depth understanding for how a massive city-scale deployment should be approached in the future. In this work, firstly we will investigate the Key Performance Indicators (KPIs) from the perspective of the first car and the RSUs, while the second vehicle acts as an interferer when being within coverage. Secondly, we will present a V2V scenario.

Three RSUs were deployed at first at three locations around the City of Bristol, UK (as shown in Fig~\ref{fig:heatmapResultsHP1}). \emph{Hydrogen-RSU} was mounted at the height of around \textasciitilde\SI{8}{\meter}, on a curvy, narrow road very close to a blind T-junction. The second one (\emph{Helium-RSU}) was installed on the wall of a building next to a straight road with some foliage at the sides at \textasciitilde\SI{5}{\meter}. Finally, \emph{Lithium-RSU} was placed on the balcony of a tall building (at \textasciitilde\SI{25}{\meter} height), next to a wide road, providing the most LOS coverage compared to the other RSUs. The different locations and buildings were chosen to evaluate how the position of a RSU can affect the performance of the network.

Two vehicles (as in Fig.~\ref{fig:car}), equipped with one OBU each, were driving randomly around the city. The second OBU unit shown in Fig.~\ref{fig:units2} was there for backup purposes only. All the devices in our system generated and transmitted a CAM per NIC every \SI{10}{\milli\second}. Each CAM, encapsulating the information described in Sec.~\ref{sub:devices}, was logged at the transmitter and the receiver side. The log files generated were used later to produce the results that will be described in the next section.

In the next section, we will present our preliminary results. We will focus our performance investigation on some meaningful KPIs related to our research and will try and comprehend the different advantages and drawbacks of our system analysing our findings. Throughout the three days of field trials, we exchanged \textasciitilde\SI{50}{\million} CAMs. Some of our results will use a subset of these exchanged messages. Our entire dataset is available for download in~\cite{dataset}. To the best of our knowledge is one the largest data repositories focused on V2X communications.

\subsection{Preliminary Results}
Firstly, we start with the V2I scenario. Fig.~\ref{fig:heatmapResults} presents the heatmap results for the PDR from all CAMs transmitted from a RSU and received at the vehicle side. The results present the PDR for the vehicle no. 1. Vehicle no. 2 acts as an interferer, as mentioned before, when both vehicles are within the same RSU coverage range. Finally, the red crosses, show the position of a vehicle when a CAM broadcast was successfully received at the RSU side. 

Figs.~\ref{fig:heatmapResultsHP1} and~\ref{fig:heatmapResultsLP1} show the PDR results when both vehicles were driving within the coverage regions of the RSUs. Figs.~\ref{fig:heatmapResultsHP2} and~\ref{fig:heatmapResultsLP2} show the results when only vehicle no. 1 was within coverage. As described, the DSRC CAMs are being broadcast from all NICs every \SI{10}{\milli\second} without having any coordination on the channel usage. As shown, there is a significant PDR difference of 30\% between the different scenarios, for both the HP and LP transceivers. This is because, the second vehicle, acting as an interferer, led to a big number of frame collisions and longer MAC-layer contention intervals at the receiver side. 

\begin{figure}[t]     
\centering
\includegraphics[width=\textwidth]{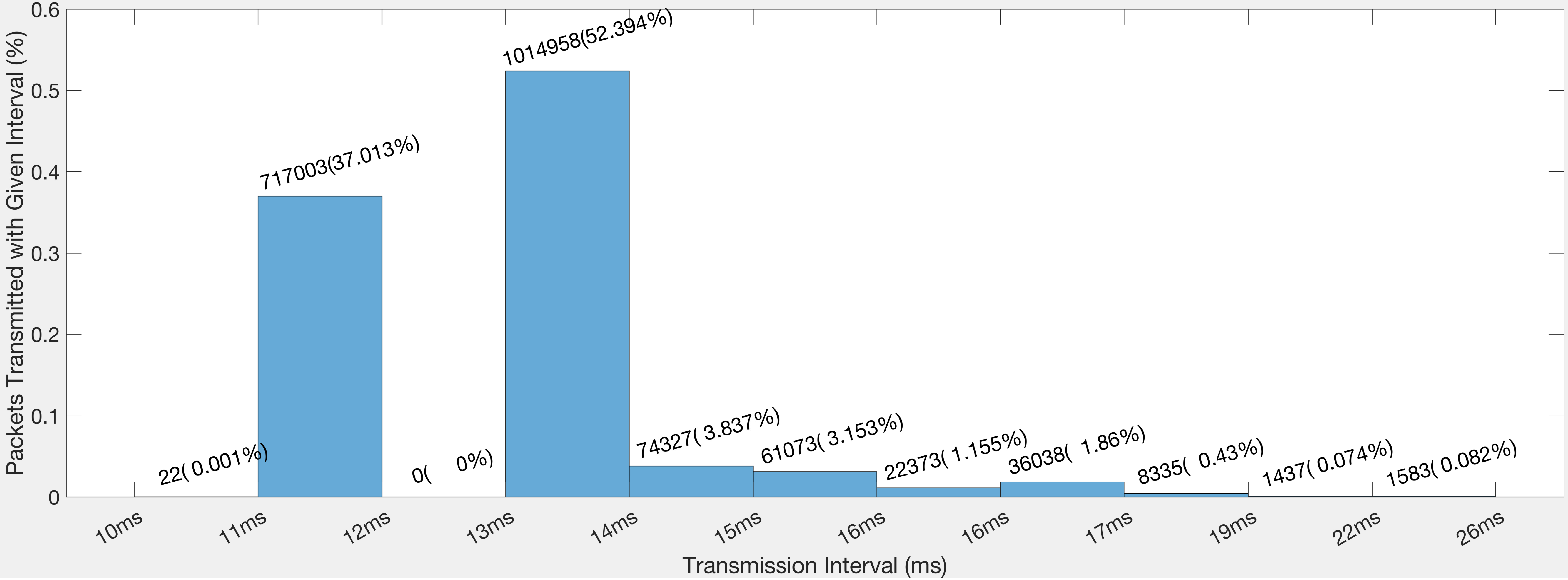}
    \caption{Transmission Intervals between two DSRC CAMs.}
    \label{fig:transmissionInterval}
\end{figure}

The difference can be observed at the RSU side as well. As shown, the heatmap data overlap with the red crosses in Figs~\ref{fig:heatmapResultsHP2} and~\ref{fig:heatmapResultsLP2}, while they do not precisely match the heatmap in the first two figures. This means that when the interfering vehicle was present, vehicle no. 1 was not always able to establish a bidirectional communication link with the RSUs.

In Fig.~\ref{fig:transmissionInterval}, we present the frequencies of the transmission interval between two DSRC CAM. This is an example from Hydrogen-RSU for all CAMs transmitted throughout one day of field trials. The remaining devices and days produced similar results, therefore will not presented in this work. As shown, even though the CAM transmission interval was set at \SI{10}{\milli\second}, our testbed generates frames at a different rate. Most of the frames are generated and exchanged either every \SI{12}{\milli\second} or \SI{14}{\milli\second}. This was expected as our devices are built upon a single-core CPU, which executes tasks with the same priority according to the Linux \emph{Deadline I/O Scheduler}. To that extent, the CPU cannot fetch/push CAMs streams towards the transceivers at a constant I/O rate. These inconsistencies should be taken into account when designing vehicular applications with strict latency requirements. Generating and processing the packets at a stronger Fog node computer, and using the transceivers as the medium to exchange the packets, will significantly improve the consistency of the transmission rate.

\begin{figure}[t]     
\centering
\includegraphics[width=\textwidth]{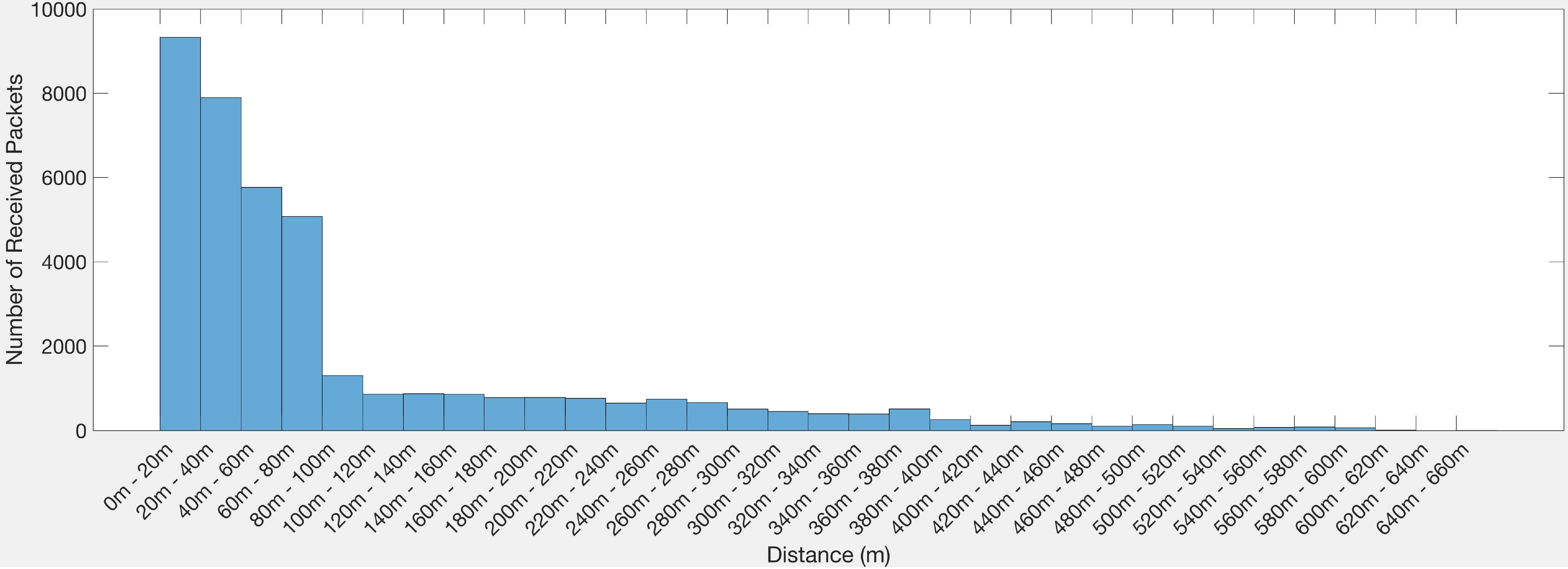}
    \caption{Awareness Horizon for the V2V Scenario - HP transceiver.}
    \label{fig:awarenessHorizon}
\end{figure}

Finally, Fig.~\ref{fig:awarenessHorizon} presents the awareness horizon for the V2V scenario, i.e. the Euclidean distance between the vehicles when a CAM is received. For this experiment, two vehicles were driving at opposing directions on a highway section of the road exchanging CAMs every time they were crossing paths. As shown in Fig.~\ref{fig:transmissionInterval} most packets are being transmitted every \SI{12}{\milli\second} or every \SI{14}{\milli\second}. Given that the vehicles drive at a constant speed, we can estimate that a similar number of packets was exchanged at every distance interval. We observe that when the vehicles are in close proximity, a bigger number of packets is being received compared to longer distances. When the vehicles are more than \SI{80}{\meter} apart, most of the packets are never delivered. Similar performance can be observed in the rural and urban trials conducted. From the above, we can observe that using the previously described setup, we can achieve adequate V2V communications for up to about \SI{80}{\meter}. For sensor features exchange at longer distances, a multi-hop communication using V2V or V2I links is necessary.

\section{Conclusions and Future Work}\label{sec:conclusions}

In this work, we presented a city-scale ITS-G5 network for next-generation ITSs. Our testbed can be used to test different networking protocols for CAVs. Utilising COTS devices can be costly and risky as their performance may be inadequate. Therefore, prototyping our testbed, we managed to reduce the deployment cost and get an in-depth understanding of the requirements and limitations of real-world large-scale deployments. The customisability and the open-source nature of our testbed are of paramount importance as different parameters can be tweaked to enhance the performance of the system and address any drawbacks.

Conducting some initial field trials, we observed the behaviour of our deployment under different conditions and scenarios. Some critical observations can be the inconsistency at the data generation, proving the necessity for a Fog computing implementation and the real-world performance evaluation that proves the need for more sophisticated MAC-layer access schemes and a centralised control plane. Our dataset of the exchanged CAMs can be downloaded from~\cite{dataset}. In the future, we intend to expand the deployed locations of our testbed, to provide almost-city-scale availability for vehicular applications. What is more, we will integrate SDN-like and Fog computing capabilities with our system to enhance its performance and scalability. Finally, a cybersecurity framework~\cite{Giovanni} will be introduced on top of our design, to secure the V2V and V2I links for potential malicious threats.

\section*{Acknowledgements}
This work was partially supported by the University of Bristol and the Engineering and Physical Sciences Research Council (EPSRC) (grant EP/I028153/1). It is also supported in part by the Innovate UK FLOURISH project under Grant no. 102582.

\bibliographystyle{splncs04}
\bibliography{bib.bib}
\end{document}